\documentclass[12pt]{article}

\usepackage{multirow}
\usepackage{comment}
\usepackage{epsfig}
\usepackage{epstopdf}
\usepackage{amssymb,amsmath}
\usepackage{comment}

\usepackage{setspace}
\newcommand{\bea}{\begin{eqnarray}}
\newcommand{\eea}{\end{eqnarray}}

\numberwithin{equation}{section}

\begin{document}
\begin{titlepage}
%\begin{flushright}
%
%\end{flushright}
%
\vspace*{10mm}
\begin{center}
\baselineskip 25pt 
{\Large\bf
%%%%%%%%%%%%%%%%%%%%%%%%%%%%%%%%%%%%%%%%%%%%%%%%%%%
SMART U(1)$_X$ $-$ Standard Model with Axion, Right handed neutrinos, Two Higgs doublets and U(1)$_X$ gauge symmetry
%%%%%%%%%%%%%%%%%%%%%%%%%%%%%%%%%%%%%%%%%%%%%%%%%%%
}
\end{center}
\vspace{5mm}
\begin{center}
{\large
Nobuchika Okada\footnote{okadan@ua.edu}, 
Digesh Raut\footnote{draut@udel.edu}, 
and Qaisar Shafi\footnote{qshafi@udel.edu}
}
\end{center}
\vspace{2mm}

\begin{center}
{\it
$^{1}$ Department of Physics and Astronomy, \\ 
University of Alabama, Tuscaloosa, Alabama 35487, USA \\
$^{2,3}$ Bartol Research Institute, Department of Physics and Astronomy, \\
 University of Delaware, Newark DE 19716, USA
}
\end{center}
\vspace{0.5cm}
%%%%%%%%%%%%%%%%%%%%%%
\begin{abstract}
%%%%%%%%%%%%%%%%%%%%%%
To address five fundamental shortcomings of the Standard Model (SM) of particle physics and cosmology,  
we propose a SMART U(1)$_X$ model which is a $U(1)_X \times U(1)_{PQ}$ extension of the SM.  
The $U(1)_X$ gauge symmetry is a generalization of the well-known $U(1)_{B-L}$ symmetry and $U(1)_{PQ}$ is the global Peccie-Quinn (PQ) symmetry. 
Three right handed neutrinos are added to cancel $U(1)_X$ related anomalies, and they play a crucial role in understanding the observed neutrino oscillations and explaining the observed baryon asymmetry in the universe via leptogenesis. 
The PQ symmetry helps resolve the strong CP problem and also provides axion as a compelling dark matter (DM) candidate. 
The $U(1)_X$ gauge symmetry 
enables us to implement the inflection-point inflation scenario with $H_{inf} \lesssim 2 \times 10^{7}$ GeV, where $H_{inf}$ is the value of Hubble parameter during inflation. 
This allows us to overcome a potential axion domain wall problem as well as the axion isocurvature problem. 
The SMART U(1)$_X$ model can be merged with $SU(5)$ as we briefly show. 
%
%Finally, it can also be compatible with the recently proposed Trans-Planckian Censorship Conjecture 
%which requires $H_{inf} < 1$ GeV, a condition that can also be realized with the IPI scenario. 
%
%%%%%%%%%%%%%%%%%%%%%%%
\end{abstract}
\end{titlepage}

\tableofcontents
\newpage

%%%%%%%%%%%%%%%%%%%%%%%%%%%%%%%%%
\section{Introduction}
\label{sec:Intro}
%%%%%%%%%%%%%%%%%%%%%%%%%%%%%%%%%
A variety of cosmological and particle physics observations have demonstrated some shortcomings of the Standard Model (SM) of particle physics. 
Cosmological and astrophysical observations strongly support, indeed require, the existence of dark matter (DM) which accounts for about 25\% of the total (critical) energy density in the universe \cite{Aghanim:2018eyx}. 
However, no viable non-baryonic DM candidate exists in the SM. 
Experimental observations of neutrino oscillations and flavor mixings indicate that neutrinos have a tiny but non-zero mass \cite{Tanabashi:2018oca}. 
But neutrinos are massless to all orders in perturbation theory in the SM.   
Cosmological and astrophysical observations have also confirmed that ordinary baryonic matter dominates over the anti-baryons in the universe \cite{Tanabashi:2018oca}. 
The SM fails to generate this so-called baryon asymmetry in the universe (BAU) \cite{Canetti:2012zc}. 
Experimental measurement of the electric dipole moment of neutron require that the effective dimensionless parameter ${\bar \theta}_{QCD}$ of the SM must be tiny, 
${\bar \theta}_{QCD} \simeq 0.7 \times 10^{-11}$ \cite{Afach:2015sja}. 
This is the so-called strong CP problem \cite{Peccei:2006as} of the SM. 
Finally, according to the current cosmological paradigm, 
the universe experienced cosmic inflation, a period of accelerated expansion in the early stages of its evolution. 
Inflation solves two major problems of standard big bang cosmology \cite{Guth:1980zm}, 
namely the origin of the observed spatial flatness of the universe and the observed uniformity of the cosmic microwave background radiation with $\delta T/T \simeq 10^{-5}$ \cite{Fixsen:2009ug}.  
Moreover, the primordial density fluctuations generated during inflation can seed these tiny fluctuations which are essential to reproduce the observed large scale structures of the universe. 
The SM needs to be extended to accommodate a realistic inflationary scenario.

It is clear that these five fundamental shortcomings of the SM are at the frontiers of high energy physics and cosmology research and crucial to understand the origin and evolution of our universe. 
Equivalently, it implies that the SM is at best an effective theory description of nature and needs to be supplemented with new physics beyond the SM. 
In this paper, we propose a gauged $U(1)_X$ extension of the SM which addresses all these shortcomings and also offer some testable predictions\footnote{See also the SMASH (Standard Model$-$axion$-$seesaw$-$Higgs portal inflation) model proposed in Ref.~\cite{Ballesteros:2016xej} to address the five fundamental shortcomings of the SM by extending the latter with the global PQ symmetry.}.

We augment the SM with $U(1)_X \times U(1)_{PQ}$ symmetry, 
where $U(1)_{PQ}$ is the well-known global Peccie-Quinn (PQ) symmetry \cite{Peccei:1977hh}, and $U(1)_X$ \cite{Appelquist:2002mw} is a generalization of the well-known $U(1)_{B-L}$ gauge symmetry \cite{mBL}. 
The generalized $U(1)_X$ charge of each field is defined as a linear combination of its hypercharge and $B-L$ charge, and is  determined by a single free parameter $x_H$ \cite{Oda:2015gna}. 
The $B-L$ charges for the particles are reproduced in the limit $x_H \to 0$. 
To cancel the $U(1)_X$ associated anomalies, 
three generations of SM singlet Majorana right handed neutrinos (RHNs) are added. 
These SM singlet RHNs explain the origin of observed neutrino masses via type-I seesaw mechanism \cite{Seesaw}.  
Furthermore, the RHNs can generate the observed BAU via leptogenesis \cite{Fukugita:1986hr}. 
In addition to a SM singlet Higgs field which breaks $U(1)_X$ symmetry, 
the model also contains a SM and $U(1)_{X}$ singlet Higgs field which breaks $U(1)_{PQ}$ symmetry
and a pair of SM doublet Higgs fields which are crucial to implement the PQ symmetry. 
In this regard, our model is a $U(1)_X$ extension of the well-known Dine-Fishler-Serednicki-Zhitnitsky (DFSZ) model \cite{Dine:1981rt, Zhitnitsky:1980tq}.

The PQ symmetry solves the strong CP problem and also provides axion as a compelling DM candidate \cite{Weinberg:1977ma}.  
The axion models can potentially encounter two major cosmological problems, namely, 
the axion domain wall problem and the isocurvature problem. 
For a review, see, for example, Ref.~\cite{Kawasaki:2013ae}. 
The topological defects (strings and domain walls) associated with PQ symmetry breaking can potentially dominate the energy density of the universe which is inconsistent with the cosmological observation. 
In addition, if inflation takes place after PQ symmetry breaking, 
it can induce isocurvature fluctuations which are severely constrained by observations \cite{Aghanim:2018eyx}. 
Both of these problems can be resolved in an inflation scenario\footnote{For a resolution of the domain wall problem without inflation, see Ref.~\cite{Kibble:1982ae}.} with a Hubble parameter during inflation $H_{inf} \lesssim 2 \times 10^{7}$ GeV. 
Such a low value for the Hubble parameter cannot be realized in a simple inflation scenario based, say, on a Coleman-Weinberg or Higgs potential with a minimal coupling to gravity \cite{Shafi:2006cs}, or a quartic potential with non-minimal coupling to gravity \cite{Okada:2010jf}; both scenarios predict $H_{inf} \simeq 10^{13-14}$ GeV \cite{Okada:2014lxa}.   
This leads us to consider the inflection-point inflation (IPI) scenario \cite{Okada:2016ssd} (see also Ref.~\cite{Ballesteros:2015noa}). 
To realize this, 
it is crucial that the inflaton field which drives inflation has both gauge and Yukawa interactions. 
The Higgs field which breaks the $U(1)_X$ gauge symmetry is a unique candidate in our model for the inflaton field. 
We will refer to this model as SMART U(1)$_X$:  SM with Axion, Right handed neutrinos, Two Higgs doublets and U(1)$_X$ gauge symmetry.

We also consider a merger of the SMART U(1)$_X$ model with grand unification. 
It was previously demonstrated in Ref.~\cite{Okada:2017dqs} that the SM quark and lepton representations in a $U(1)_X$ extended SM can be embedded inside $SU(5)$ for a fixed $x_H = -4/5$. 
Therefore this grand unified theory (GUT) scenario, the $U(1)_X$ symmetry explains the origin of charge quantization. 
By incorporating the PQ symmetry, 
the model we consider is $SU(5)\times U(1)_X \times U(1)_{PQ}$. 
With the addition of suitable new vector-like fermions, 
the three SM gauge couplings successfully unify at $M_{GUT} \simeq 9.8 \times 10^{15}$ GeV. 
These new fermions are also essential  for a successful implementation of the IPI scenario, 
which plays an important role in solving both the axion domain wall and isocurvature problems. 
In addition, we show that the SM Higgs potential can also be stabilized in the presence of the new vector-like fermions.

The recently proposed Trans-Planckian Censorship Conjecture (TCC) \cite{Bedroya:2019snp}, when applied to slow-roll inflation, significantly lowers the bound on the value of Hubble parameter during inflation, $H_{inf} \lesssim 1$ GeV \cite{Bedroya:2019tba}. 
An IPI inflation scenario consistent with TCC was recently examined by the authors of this paper \cite{Okada:2019yne} in a model whose particle content matches that of the SMART U(1)$_X$ model with $x_H =0$. 
Therefore, the model in Ref.~\cite{Okada:2019yne} is the $B-L$ version of the SMART U(1)$_X$. 
%We can generalize it by promoting the $U(1)_{B-L}$ symmetry to the $U(1)_X$ symmetry, 
%
%Here, we will generalize by promoting the $U(1)_{B-L}$ symmetry to the $U(1)_X$ symmetry, 
%and as an example, we will consider an $SU(5)$ GUT extension of the SMART U(1)$_X$ which is consistent with the TCC conjecture. 

This paper is organized as follows: 
In Sec.~2 we describe the SMART U(1)$_X$ model. 
We discuss the axion DM scenario in Sec.~3 and show that both the axion domain wall and isocurvature problems can be solved if $H_{inf} \lesssim 2 \times 10^{7}$ GeV. 
An IPI scenario with this constraint is discussed in Sec.~4, 
where we consider reheating after inflation and identify the model parameters required for successful leptogenesis. 
In Sec.~5 we consider the merger of $SU(5)$ with the SMART U(1)$_X$.
% and in Sec.~6 we discuss a GUT extended  SMART U(1)$_X$ which is consistent with the TCC conjecture. 
Our results are summarized in Sec.~6.

%%%%%%%%%%%%%%%%%%%%%%%%%%%%%%%%%%%%%
\begin{table}[t]
\begin{center}
\begin{tabular}{|c|ccc|c|c|c|}
\hline
                       &SU(3)$_C$        &SU(2)$_L$       &U(1)$_Y$          & U(1)$_{X}$                        & U(1)$_{PQ}$
\\ 
\hline
$q^{i}$           &{\bf 3 }             &{\bf 2}            &$\;\;1/6$          & $(1/6) x_{H} + (1/3)$         & $\;\;1  $   
\\
$(u^{c})^i$     &${\bf 3 }^*$     &{\bf 1}           &$-2/3$              & $(-2/3) x_{H} + (-1/3)$       & $\;\;1 $   
\\
$(d^{c})^i$     & ${\bf 3 }^*$    &{\bf 1}           &$\;\;1/3$          & $(+1/3) x_{H} + (-1/3)$      & $\;\;1$   
\\
\hline
$\ell^{i}$         &{\bf 1 }            &{\bf 2}            &$-1/2$              & $(-1/2) x_{H} + (-1)$           &$\;\;1$
\\
$(e^{c})^i$      &{\bf 1 }            &{\bf 1}           &$\;\;1$              & $(+1) x_{H} + (+1)$           &$\;\;1$   
\\
$(N^{c})^{i}$   &{\bf 1 }           &{\bf 1}            &$\;\;0$             & $ (+1)$                                & $\;\;1$   
\\
\hline
$H_u$                & {\bf 1 }          &{\bf 2}            &$1/2$              & $(+1/2) x_{H}$                    & $-2$   
\\  
$H_d$                & {\bf 1 }          &{\bf 2}            &$- 1/2$            & $(-1/2) x_{H}$                     & $-2$   
\\  
$\Phi$                &{\bf 1 }           &{\bf 1}            &$\;\;0$             &$(-2) $                                  & $-2$  
\\
$S$                    & {\bf 1 }           &{\bf 1}            &$\;\;0$            &$\;\;0$                                  & $\;\;4$   
\\
\hline
\end{tabular}
\end{center}
\caption{
Particle content of the SMART U(1)$_X$ model where all  fermions fields are left-handed. 
In addition to the three generations of SM fermions ($i=1,2,3$), we have three Majorana neutrinos, $(N^c)^{i}$, and the scalar sector has two SM doublet Higgs, $H_{u,d}$, and two SM singlet Higgs, $\Phi$ and $S$. 
}
\label{tab:1}
\end{table}
%%%%%%%%%%%%%%%%%%%%%%%%%%%%%%%%%%%%%%%%%%%%%%%

%%%%%%%%%%%%%%%%%%%%%%%%%%%%%%%%%%%%%%
\section{SMART U(1)$_X$}
\label{sec:model}
%%%%%%%%%%%%%%%%%%%%%%%%%%%%%%%%%%%%%%
The particle content of our model is listed in Table~\ref{tab:1}, where all fermion fields are left-handed. 
The $U(1)_X$ charge of each particle is defined as $Q_X = x_{H} Q_{Y}  + x_{\Phi} Q_{B-L}$, 
where $Q_Y$ and $Q_{B-L}$ are their SM hypercharge and $B-L$ (baryon minus lepton) number, respectively. 
We fix $x_\Phi = 1$ without loss of generality so that the $U(1)_X$ charges of all the particles are uniquely determined by a single free parameter $x_H$. 
The $B-L$ charges are reproduced in the limit $x_H \to 0$. 
In addition to three generations of SM quarks and leptons ($i = 1,2,3$), 
there are three SM singlet Majorana neutrinos, $(N^c)^{i}$,  which cancel all the $U(1)_X$ related gauge and mixed gauge-gravitational anomalies.  
The scalar sector includes four complex scalar fields, namely the two SM doublet Higgs fields, $H_{u,d}$, and two SM singlet Higgs fields, $\Phi$ and $S$.

The gauge $U(1)_X$ and global $U(1)_{PQ}$ invariant Higgs potential is given by 
\bea  
V &=& 
 - \sum_{i = u,d} \mu_i^2 \left( H_i^{\dagger}H_i \right)
+\sum_{i = u,d}\lambda_i  \left( H_i^{\dagger}H_i \right)^2 
+  \left( \sqrt{2}\Lambda_{s}\left( H_u \cdot H_d \right) S + {\rm h. c.}\right)
\nonumber \\
&&
+\lambda_{\phi} \left(\Phi^\dagger \Phi  - \frac{v_{BL}^2}{2}\right)^2 
+\lambda_{S} \left(S^\dagger S  - \frac{v_{PQ}^2}{2}\right)^2 
+ {\rm mixed \; quartic \; terms}, 
\label{eq:HPot}
\eea
where the couplings parameters are all chosen to be positive, the {\it dot} represents contraction of $SU(2)$ indices by {\it epsilon} tensor,  
and the vacuum expectation values (VEVs) of $\Phi$ and $S$ are given by $ \langle \Phi \rangle =  v_X/\sqrt{2}$
and $ \langle S \rangle =  v_{PQ}/\sqrt{2}$, respectively. 
The mixed quartic interactions includes terms such as $(H_i^\dagger H_i)(\Phi^\dagger \Phi)$, $(H_i^\dagger H_i)(S^\dagger S)$, etc.  
For simplicity, we assume these mixed quartic couplings to be adequately small, and they do not alter our results. 
We also assume that 
both $U(1)_X$ and $U(1)_{PQ}$ 
symmetry are broken at a scale much higher than the electroweak scale as well as $\Lambda_s$.
%, such that the SM singlet Higgs sector decouples at low energies. 

We parameterize $S$ and $\Phi$ in terms of real fields as follows:   
\bea
\Phi(x) &=& \frac{1}{\sqrt{2}}\left(\phi(x) + v_{X}\right) e^{i \chi(x)/ v_{X}}, 
\nonumber \\
S(x) &=& \frac{1}{\sqrt{2}}\left(s(x) + v_{PQ}\right) e^{i a(x)/ v_{PQ}}. 
\label{eq:phiABH} 
\eea 
Note that the Higgs field $\Phi$ is charged under both $U(1)_X$ and $U(1)_{PQ}$, 
whereas $S$ is charged only under $U(1)_{PQ}$. 
After $S$ and $\Phi$ fields acquire their VEVs, 
the would-be Nambu-Goldstone (NG) boson $\chi (x)$ is absorbed by the $U(1)_X$ gauge boson ($Z^\prime$) 
and $a(x)$ is identified with the axion. 
%\footnote{The QCD interactions generate non-zero mass for the axion after the QCD phase transition, see, for example, \cite{Borsanyi:2016ksw} for latest lattice calculation results.} 
After the $U(1)_X \times U(1)_{PQ}$ symmetry breaking, 
the masses of $\phi$, $s$ and $Z^\prime$ gauge boson are respectively given by    
\bea 
m_\phi = \sqrt{2 \lambda_\phi} v_{X}, \; \; 
m_s = \sqrt{2 \lambda_s} v_{PQ},  \;\;  
m_{Z^\prime} = 2 g v_{X},
\label{eq:mass}
\eea
where $g$ is the $U(1)_X$ gauge coupling.

Our assumption of negligible mixed quartic couplings allows us to separately analyze the SM doublet Higgs potential at low energies from the singlet Higgs sector. 
In Eq.~(\ref{eq:HPot}), the PQ symmetry breaking Higgs field $S$ generates the mixing mass term between the two doublet Higgs fields, 
$m_{mix}^2 (H_u. H_d)$, 
where $m_{mix}^2 = \Lambda_{s} v_{PQ}$.  
Because of their charge assignments, 
$H_u$ ($H_d$) only couple with up-type  (down-type) SM fermions (see Eqs.~(\ref{eq:LY1}) and (\ref{eq:LY2}) below). 
The doublet Higgs potential at low energies in this case is the same as that of the type-II two Higgs doublet extension of the SM. 
Since the two Higgs doublet model is well studied in the literature \cite{Branco:2011iw}, 
we will skip the detailed phenomenology of the Higgs potential at low energies.

In addition to the Yukawa interactions for the SM quarks, 
\bea
{\cal L}  \supset 
\sum_{i,j=1}^{3} Y_u^{ij} \left(q^i . H_u\right) (u^{c})^j 
+ \sum_{i,j=1}^{3} Y_d^{ij} \left(q^{i}. H_d\right) (d^{c})^j, 
\label{eq:LY1}
\eea 
the Lagrangian includes the following new Yukawa interactions involving the Majorana neutrinos: 
\bea  
   {\cal L} \supset \sum_{i,j=1}^{3}Y_\ell^{ij}  \left(\ell^{i}. H_d\right) (e^{c})^j  
-\frac{1}{2} \sum_{i,j=1}^{3} Y_D^{ij}\left( \ell^{i}. H_u\right) \left(N^c\right)^{j} 
- \left(\frac{1}{2} \sum_{i=1}^{3} Y_i  \Phi {\left(N^c\right)}^{i} \left(N^c\right)^{i}  +{\rm h.c.}\right),
\label{eq:LY2}
\eea 
where $Y_D$ ($Y_i$) is the Dirac (Majorana) Yukawa coupling, 
and we have chosen a flavor-diagonal basis for $Y_i$. 
After the $U(1)_X$ and electroweak symmetry breakings,  
the Dirac and the Majorana masses for the neutrinos are generated, 
\bea 
m_D^{ij}= \frac{Y_D^{ij}}{\sqrt{2}} v_{u}, \; \; 
m_{N^i}= \frac{1}{\sqrt{2}} Y_i  v_{X}, 
\label{eq:masses}
\eea
where $ \langle H_u^0 \rangle = v_u/\sqrt{2}$ is the VEV of the charge neutral component of $H_u$.

%%%%%%%%%%%%%%%%%%%%
\section{Axion Dark Matter}
\label{sec:axionDM}
%%%%%%%%%%%%%%%%%%%% 
As discussed before, the axion DM scenario possibly suffers from two major cosmological problems, namely, 
the axion domain wall problem and the axion DM isocurvature problem. 
This domain wall problem can be solved if inflation takes places after the PQ symmetry breaking, 
or equivalently, $H_{inf} < F_a=v_{PQ}/N_{DW}$, 
  where $F_a$ is the axion decay constant and $N_{DW}$ is the domain wall number. 
In our case, $N_{DW} = 12$.
The measurement of supernova SN 1987A pulse duration provides a model-independent constraint
 on the axion decay constant $F_a  \gtrsim 4 \times 10^8$ GeV \cite{Raffelt:2006cw}.  
We will shortly show that the resolution of the isocurvature problem leads to an even stronger constraint on $H_{inf}$ than the one required to solve the axion domain wall problem, $H_{inf} < F_a$.

%%%%%%%%%%%%%%%%%%%%
\subsection{Relic Abundance of Axion Dark Matter}
\label{subsec:axionDM}
%%%%%%%%%%%%%%%%%%%% 
The PQ symmetry breaking produces cosmic strings, which can efficiently decay into axions \cite{Kawasaki:2013ae}. 
Because inflation takes place after the PQ symmetry breaking in our setup, 
the axions produced from cosmic string decay contribute negligibly to the relic abundance. 
Although QCD interactions can also produce axions in the thermal plasma, 
their relic abundance from such processes was shown to be negligible compared to the observed abundance of DM in Ref.~\cite{Graf:2010tv}. 
At the end of QCD phase transition, 
the coherently oscillating axion field behaves like cold DM and contribute dominantly to the relic abundance of the axion \cite{Kawasaki:2013ae},   
\bea
\Omega_a^{\rm mis} h^2 \simeq  0.18 \; \left( \theta_m^2 
+\delta{\theta_m}^2\right)\; \left(\frac{F_a}{10^{12}\; {\rm GeV}}\right)^{1.19} 
\simeq 0.18 \; \theta_m^2  \left(\frac{F_a}{10^{12}\; {\rm GeV}}\right)^{1.19} . 
\label{eq:DMab}
\eea
Here, the misalignment angle $\theta_m$ is the initial displacement of axion field from the potential minima at the onset of oscillations,  
and $\delta{\theta_m} = H_{inf}/(2\pi F_a)$ is the fluctuation of $\theta_m$ generated by inflation. 
A natural choice for the misalignment angle is $\theta_a \simeq 1$, 
and we have used $H_{inf} <  F_a$ to obtain the final expression in  Eq.~(\ref{eq:DMab}).  
Requiring that the axion fully account for the DM in the universe, 
$\Omega_{a} h^2 = 0.120 \pm 0.0012$ \cite{Aghanim:2018eyx}, 
the axion decay constant $F_a$ is determined as a function of $\theta_m$, 
\bea
F_a \simeq  7.11 \times 10^{11} \; {\rm GeV} \; \theta_m^{-1.68}.  
\label{eq:theta}
\eea

%%%%%%%%%%%%%%%%%%%%%%%%%%%%%%%%%%%%%%
\subsection{Axion Dark Matter Isocurvature Fluctuations}
%%%%%%%%%%%%%%%%%%%%%%%%%%%%%%%%%%%%%%
If inflation takes place after the PQ symmetry breaking, 
it induces isocurvature fluctuation in the axion DM power spectrum \cite{Aghanim:2018eyx},  
\bea
{\cal P}_{\rm iso} =\left(\frac{H_{inf}}{\pi \theta_m  F_a }\right)^2,  
\eea
which is constrained by Planck measurements \cite{Aghanim:2018eyx}, namely
\bea
\beta_{\rm iso} \equiv \frac{{\cal P}_{\rm iso}(k_*)}{{\cal P}_{\rm iso}(k_*)+ {\cal P}_{\rm adi}(k_*)} < 0.038, 
\label{eq:isoPlanck}
\eea
where ${\cal P}_{\rm adi} (k_*) \simeq 2.2 \times 10^{-9}$ \cite{Aghanim:2018eyx} is the adiabatic power spectrum and $k_* = 0.05$ Mpc$^{-1}$ is the pivot scale. 
Combining Eqs.~(\ref{eq:theta}) and (\ref{eq:isoPlanck}) yields an upper bound on $H_{inf}$,  
\bea
H_{inf} < 2.08 \times 10^7 \;  {\rm GeV} \left(\frac{F_a}{7.11 \times 10^{11}\; {\rm GeV}}\right)^{0.405}.  
\label{eq:Hiso1}
\eea
Together with Eqs.~(\ref{eq:theta}), 
we find that $\delta \theta_m^2/ \theta_m^2 \simeq 10^{-11}$, 
or equivalently, 
$H_{inf}/{F_a} \lesssim 3.0\times 10^{-5}\; \theta_m$. 
This shows that the axion DM isocurvature constraint imposes  a much stronger restriction on $H_{inf}$ than the one required to solve the axion domain wall problem,  $H_{inf} < F_a$.

Let us set $\theta_m$ to be ${\cal O} (1)$ and $F_a  = 7.11 \times 10^{11}$ GeV such that the axion saturates the observed DM abundance. 
We therefore obtain an upper bound on the Hubble parameter during inflation,  $H_{inf} < 2.08 \times 10^{7}$ GeV.  
In the next section, 
we discuss an IPI scenario with the Hubble parameter value of this magnitude which also solves both the axion domain wall and axion DM isocurvature problems.

%%%%%%%%%%%%%%%%%%%%%%%%%%%%%%%%%%%%%%
\section{Inflection-Point Inflation}
%%%%%%%%%%%%%%%%%%%%%%%%%%%%%%%%%%%%%%
Let us begin by highlighting the key results of the inflection-point inflation (IPI) scenario driven by a real scalar field $\phi$. 
See Ref.~\cite{Okada:2016ssd} for details.   
An inflaton potential which exhibits an approximate inflection-point around $\phi = M$ can be expressed as   
\bea
V(\phi)\simeq V_0 +\ V_1 (\phi-M) +  \frac{V_2}{2} (\phi-M)^2 + \frac{V_3}{6} (\phi-M)^3, 
\label{eq:PExp}
\eea
   where $V_0 = V(M)$, $V_n \equiv  {\rm d}^{n}V/{\rm d} \phi^n |_{\phi =M}$, 
and we identify $\phi = M$ to be the horizon exit scale that corresponds to the pivot scale $k_* = 0.05$ Mpc$^{-1}$ used in Planck measurements \cite{Akrami:2018odb}.

With the potential in Eq.~(\ref{eq:PExp}), the inflationary slow-roll parameters are given by
\bea
  \epsilon \simeq \frac{M_P^2}{2} \left( \frac{V_1}{V_0}\right)^2, 
\; \;\;
  \eta \simeq M_P^2 \left( \frac{V_2}{V_0}\right), \; \;\;
  \zeta \simeq M_P^4 \frac{V_1 V_3}{V_0^2} ,
\eea
where $M_P = 2.43\times 10^{18}$ GeV is the reduced Planck mass. 
In terms of the slow-roll parameters, 
the inflationary predictions for the scalar spectral index ($n_s$), the tensor-to-scalar ratio ($r$), and the running of the spectral index ($\alpha$) are respectively given by 
\bea
n_s=1- 6\epsilon +2 \eta, \; 
r=16 \epsilon, \;
\alpha= 16 \epsilon \eta -24 \epsilon^2 -2 \zeta^2.  
\label{eq:INFp}
\eea     
The amplitude of the curvature perturbation $\Delta_{\mathcal{R}}^2$ is expressed as  
\bea
  \Delta_{\mathcal{R}}^2 \simeq \frac{1}{24 \pi^2} \frac{V_0}{ M_P^4 \;\epsilon}. 
\eea

The central values from the Planck 2018 results \cite{Akrami:2018odb}, 
$\Delta_{\mathcal{R}}^2= 2.195 \times 10^{-9}$ and 
 $n_s = 0.9649$, 
can be used to express $V_{1,2}$ in terms of $V_0$ and $M$:    
\bea
\frac{V_1}{M^3}&\simeq& 1.96 \times 10^3 \left(\frac{M}{M_P}\right)^3\left(\frac{V_0}{M^4}\right)^{3/2}, \nonumber \\
%\frac{V_2}{M^2}&\simeq& -1.76 \times 10^{-2} \left(\frac{1-n_s}{1-0.9649} \right) \left(\frac{M}{M_P} \right)^2 
%   \left(\frac{V_0}{M^4}\right), \nonumber \\
\frac{V_2}{M^2}&\simeq& -1.76 \times 10^{-2}  \left(\frac{M}{M_P} \right)^2 \left(\frac{V_0}{M^4}\right).
%\frac{V_3}{M} &\simeq& 6.99 \times 10^{-7} \; 
  % \left( \frac{M}{M_P} \right) \left( \frac{V_0}{M^4}   \right)^{1/2}, 
\label{eq:FEq-V1V2}
\eea
The following expression for the number of e-folds has been derived in Ref.~\cite{Okada:2016ssd}: 
\bea 
N = \frac{1}{M_P^2} \int_{\phi_e}^M \, d\phi \, \frac{V}{(dV/d\phi)} \simeq \pi \frac{V_0}{M_P^2 \sqrt{2V_1V_3}}. 
\label{eq:N}
\eea
Combining Eqs.~(\ref{eq:FEq-V1V2}) and (\ref{eq:N}) we obtain 
\bea
\frac{V_3}{M} &\simeq& 6.99 \times 10^{-7} \; 
  \left( \frac{60}{N} \right)^2\; \left( \frac{M}{M_P} \right) \left( \frac{V_0}{M^4}   \right)^{1/2}. 
\label{eq:FEq-V3}
\eea
In the following analysis, we set $N=60$ to solve the horizon problem of big bang cosmology.  
Using 
Eqs.~(\ref{eq:INFp}), (\ref{eq:FEq-V1V2}), and (\ref{eq:FEq-V3}), 
the prediction for the running of spectral index is obtained as 
$\alpha = -2.74 \times 10^{-3}$. 
This is consistent with the Planck 2018 measurements, 
$\alpha = - 0.0045\pm 0.0067$, and can be tested in the future \cite{Abazajian:2013vfg}.

Next we identify the inflaton field used in the IPI analysis with the real component of the $U(1)_X$ Higgs field, $\phi = \sqrt{2} {\rm Re} [\Phi]$.  
The inflaton potential in Eq.~(\ref{eq:PExp}) 
is identified with the renormalization group improved $U(1)_X$ Higgs potential, 
\bea
V( \phi) =  \lambda_\phi (\phi) \left( \Phi^\dagger \Phi - \frac{v_X^2}{2}  \right)^2 \simeq \frac{1}{4} \lambda_\phi ( \phi)\; \phi^4. 
\label{eq:VEff1}
\eea
To obtain the second expression, 
we have assumed that during inflation $ \phi \gg v_{PQ}$ and $\lambda_\phi (\phi)$ is determined by solving the the following renormalization group equations (RGEs):  
\bea
 \phi  \frac{d g}{d  \phi} &=& \frac{1}{16 \pi^2}  \left(\frac{72 + 64 x_H + 41 x_H^2}{6}\right) g^3,         \nonumber\\
 \phi \frac{d Y_{i}}{d  \phi}   &=& \frac{1}{16 \pi^2}\left(Y_i^2+\frac{1}{2} \sum_{j=1}^3  Y_j^2-6 g^2 \right) Y_i,   
\nonumber\\
 \phi \frac{d \lambda_\phi}{d  \phi}  &=& \beta_{\lambda_\phi}. 
  \label{eq:RGEs}
\eea
Here, the beta-function of ${\lambda_\phi}$ and is given by
\bea
\beta_{\lambda_\phi} \!=\! \frac{1}{16 \pi^2}\! \left(\!20 \lambda_\phi^2  \!- 48\lambda_\phi  g^2\!+2 \lambda_\phi\sum_{i=1}^3 Y_i^2 
%\right. 
%\nonumber \\
%&& \left.
+96 g^4 - \sum_{i=1}^3 Y_i^4\!\right).
\label{eq:BGen}
\eea

The renormalization group improved Higgs potential together with the RGE for $\lambda_\phi$ can be used to express $V_{1,2,3}$ in Eq.~(\ref{eq:PExp}) as 
\bea
\frac{V_1}{M^3}&=& \left.\frac{1}{4} (4 \lambda_\phi + \beta_{\lambda_\phi})\right|_{ \phi= M},\nonumber \\
\frac{V_2}{M^2}&=&  \left.\frac{1}{4} (12\lambda_\phi + 7\beta_{\lambda_\phi}+M \beta_{\lambda_\phi}^\prime)\right|_{ \phi= M}, \nonumber \\
\frac{V_3}{M}&=&  \left.\frac{1}{4} (24\lambda_\phi + 26\beta_{\lambda_\phi}+10M \beta_{\lambda_\phi}^\prime+M^2 \beta_{\lambda_\phi}^{\prime\prime})\right|_{ \phi= M}, 
\label{eq:ICons2}
\eea
where the prime denotes derivatives with respect to $\phi$.
To realize an approximate inflection-point at $M$, we impose $V_1/M^3\simeq 0$ and $V_2/M^2\simeq 0$, which yields  the following relations: $\beta_{\lambda_\phi} (M)\simeq -4\lambda_\phi(M)$ and $M\beta_{\lambda_\phi}^{\prime}(M) \simeq  16 \lambda_\phi (M)$. 
Assuming $g, Y_i, \lambda_\phi \ll 1$, 
we can approximate $M^2 \beta_{\lambda_\phi}^{\prime\prime}(M) \simeq - M \beta_{\lambda_\phi}^{\prime}(M) \simeq -16 \lambda_\phi(M)$, 
where we have neglected higher order terms proportional to  $g^8$, $Y_i^8$, $\lambda_\phi^4$, etc.
Thus last expression in Eq.~(\ref{eq:ICons2}) is simplified to $V_3/M \simeq 16 \;\lambda_\phi(M)$. 
Comparison with the expression for $V_3/M$ in Eq.~(\ref{eq:FEq-V3}), with $V_0 \simeq (1/4) \lambda_\phi(M) M^4$, 
leads to the following expression for $\lambda_\phi$ at the inflation scale $\phi = M$: 
\bea
\lambda_\phi(M)\simeq 4.77 \times 10^{-16} \left(\frac{M}{M_{P}}\right)^2. 
\label{eq:FEq} 
\eea
Using this expression for $\lambda_\phi(M)$, 
both the tensor-to-scalar ratio ($r$) in Eq.~(\ref{eq:INFp}) 
and the Hubble parameter during the inflation ($H_{inf}$) are uniquely determined as a function of only the single parameter, 
\bea
r &\simeq& 3.7 \times 10^{-9}  \left(\frac{M}{M_{P}}\right)^6,
\nonumber \\
H_{inf} &\simeq& \sqrt{\frac{V_0}{M_P^4}} \simeq 1.5\times 10^{10} \;{\rm GeV} \;\left(\frac{M}{M_P}\right)^3. 
\label{eq:FEqR} 
\eea
Recall that we obtained an upper bound on $H_{inf}$ in Eq.~(\ref{eq:Hiso1}) to solve the axion domain wall and isocurvature problems. 
Combined with Eq.~(\ref{eq:FEqR}),  
this leads to an upper-bound on $M$: 
\bea
\frac{M}{M_P} < 0.11 \left(\frac{F_a}{7.11 \times 10^{11} \;{\rm GeV}}\right)^{0.135}.  
\label{eq:MBound}
\eea
For the benchmark values $\theta_m \approx 1$  and $F_a  = 7.11 \times 10^{11}$ GeV, 
we obtain an upper bound $M < 0.11 \; M_P$. 
Therefore, the tensor-to-scalar ratio is predicted to be be tiny, $r < 6.55 \times 10^{-15}$, well below the reach of any foreseeable experiments.

%We note that the following results obtained below is also valid for $m_{Z^\prime}/m_{N^1}> 10$. 

%%%%%%%%%%%%%%%%%%%%%%%%%%%%%%%%%%%
\subsection{Low Energy Predictions}
\label{sec:lowenergy}
%%%%%%%%%%%%%%%%%%%%%%%%%%%%%%%%%%
In the following analysis, we will estimate the low energy values for the gauge coupling ($g$), 
the Yukawa couplings ($Y_i$), and the quartic coupling coupling ($\lambda_\phi$), 
to determine the masses of the gauge boson $Z^\prime$, Majorana neutrinos, and the inflaton. 
Let us fix the mass ratio between $Z^\prime$ and $(N^c)^1$ at the inflation scale $\phi = M$ to be  $m_{Z^\prime}/m_{N^1} = 10$, 
or equivalently, 
$Y_1 (M)/g(M) = \sqrt{2}/5$. 
The two remaining Yukawa couplings are assumed to be degenerate, $Y_{2,3} (M) = Y$. 
As we have obtained in Eq.~(\ref{eq:FEq}), IPI requires $\lambda_\phi(M)$ to be extremely small.  
Assuming $g,  Y_{1,2,3}  \gg \lambda_\phi$, 
the condition $\beta_{\lambda_\phi}(M) \simeq 0$ for the inflection-point leads to 
\bea
Y(M)\simeq 2.63\;g(M). 
\label{eq:FEq3}
\eea
Explicitly evaluating another inflection-point condition, $M\beta_{\lambda_\phi}^{\prime}(M) \simeq  16 \lambda_\phi (M)$, 
using the RGEs in Eq.~(\ref{eq:RGEs}) and the relation in Eq.~(\ref{eq:FEq3}), 
we obtain
\bea
\lambda_\phi(M) \simeq 3.95\times 10^{-5} \left(100+260 x_H+166  {x_H}^2\right) \, g(M)^6. 
\label{eq:LandG}
\eea
Substituting this $\lambda_\phi(M)$ to Eq.~(\ref{eq:FEq}), 
the gauge coupling at the inflation scale $\phi = M$ is given by 
\bea
g(M)\simeq  \frac{ 1.51\times 10^{-2}}{\left(100+260 x_H+166 {x_H}^2\right)^{1/6}}  
 \left(\frac{M}{M_{P}}\right)^{1/3}.
\label{eq:FEq2} 
\eea

Since $g(M) , Y_i(M) \ll1$, 
their low energy values are well approximated by their values at $\phi=M$, 
$g(v_X) \simeq g(M) $ and $  Y_i (v_X) \simeq Y_i (M)$.  
The low energy value of $\lambda_\phi$ is then estimated to be \cite{Okada:2016ssd}
\bea
\lambda_\phi( \phi) &\simeq& 8 \lambda_\phi(M) \left(\ln \left[\frac{ \phi}{M}\right] \right)^2 
\nonumber \\
&=&  
3.81 \times 10^{-15}\left(\frac{M}{M_P}\right)^2\left(\ln \left[\frac{ \phi}{M}\right] \right)^2, 
\label{eq:Lmu} 
\eea
where we have used Eq.~(\ref{eq:FEq}) to obtain the final expression. 
The masses of the particles in Eq.~(\ref{eq:mass}), evaluated at $\phi =v_X$, are then given by 
\bea
m_\phi &\simeq& 8.72 \times 10^{-8} v_X \left|{\rm ln}\left[\frac{v_X}{M}\right]\right|\left(\frac{M}{M_P}\right), 
\nonumber \\
m_{Z^\prime} &\simeq& \frac{ 3.02\times 10^{-2} v_X}{\left(100+260 x_H+166  {x_H}^2\right)^{1/6}} 
 \left(\frac{M}{M_{P}}\right)^{\!\!1/3}, 
\nonumber \\
m_{N^{1}} &\simeq& \frac{m_{Z^\prime}}{10}, 
\nonumber \\
m_{N^{2,3}} &\simeq& 0.93 \; m_{Z^\prime}. 
\label{eq:ratiophiz} 
\eea

%%%%%%%%%%%%%%%%%%%%%%%%%%%%%%%%
\begin{figure}[t]
\begin{center}
\includegraphics[scale=0.6]{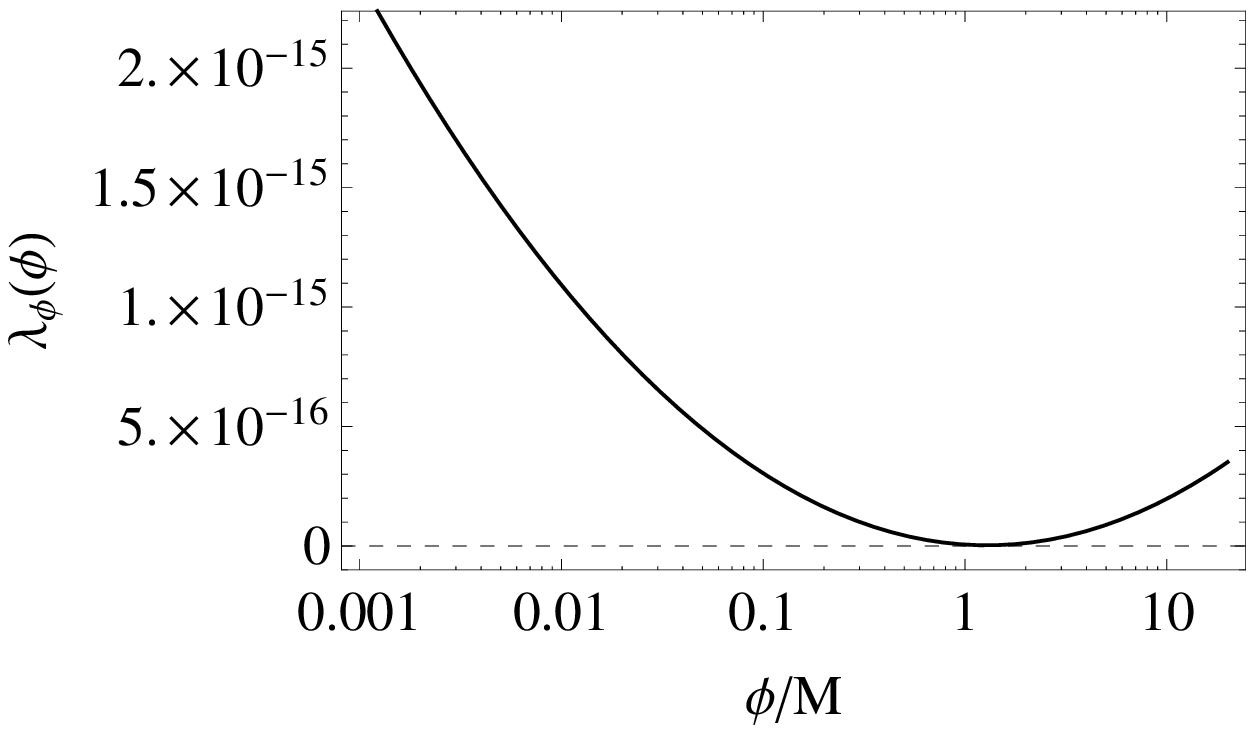} \;
\includegraphics[scale=0.6]{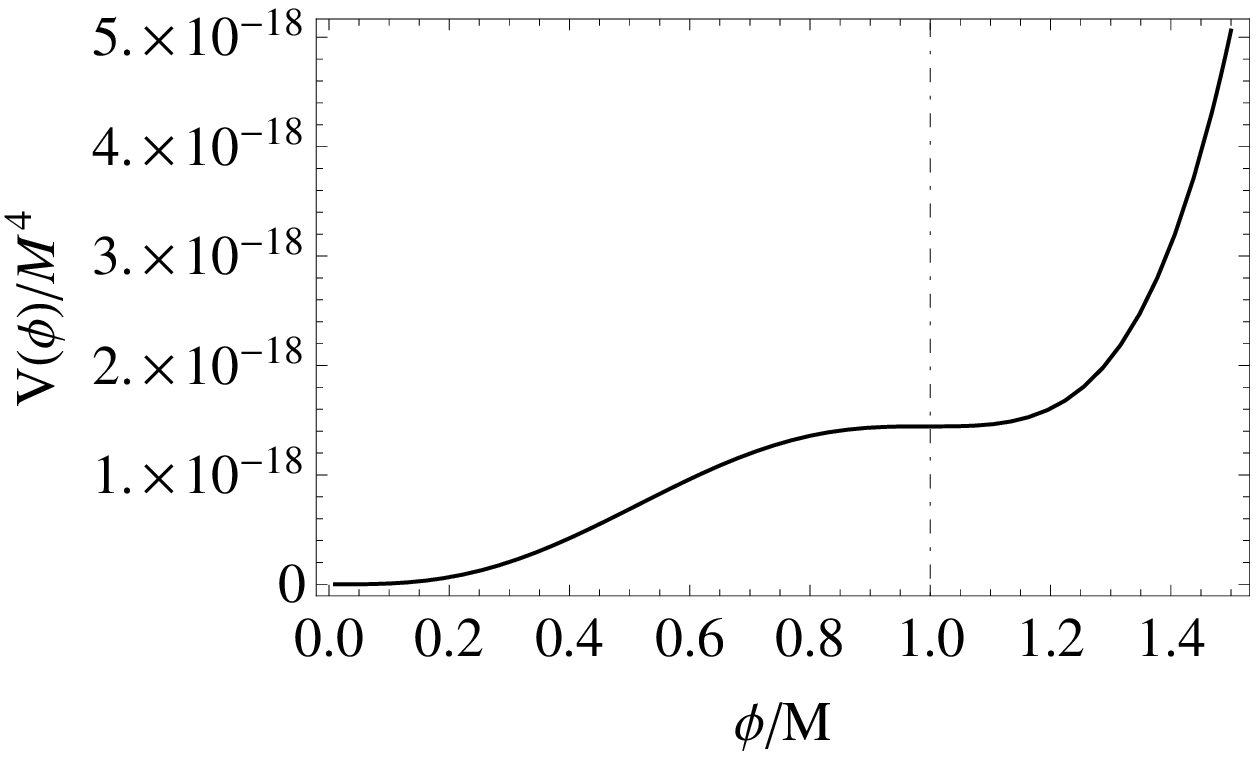}       
\end{center}
\caption{
With $M = 0.11 M_{P} $, the left panel shows the RG running of the inflaton quartic coupling as a function of $ \phi/M$, where the dashed horizontal line corresponds to $\lambda_\phi=0$. 
The right panel shows the RG improved effective inflaton potential with an approximate inflection-point at $ \phi \simeq M$ (vertical dashed-dotted line). 
}
\label{fig:InfPot}
\end{figure}
%%%%%%%%%%%%%%%%%%%%%%%%%%%%%%%%%%

In Fig.~\ref{fig:InfPot}, we plot the RG running of  the inflaton quartic coupling (left) and the corresponding RG improved effective inflaton potential (right) which exhibits an approximate inflection-point at $\phi \simeq M$ (vertical dashed-dotted line). 
The dashed horizontal line in the left panel depicts $\lambda_\phi=0$. 
Here, we have set $M = 0.11 M_{P}$ and $x_H = 0$ such that $g (M) \simeq 3.37 \times 10^{-3}$, $Y (M) \simeq 8.87 \times 10^{-3}$, and $\lambda_\phi (M) \simeq 5.77 \times10^{-18}$. 
The left panel shows that the running quartic coupling (solid curve) exhibits a minimum with almost vanishing value near $ \phi \simeq M$, namely, 
$\lambda_\phi (M)\simeq 0$ and $\beta_{\lambda_\phi} (M) \simeq 0$.\footnote{To stabilize the inflaton potential, similar conditions have been obtained in Ref.~\cite{Okada:2015lia}.} 
This behavior for the RG running of $\lambda_\phi$ is a key to realize an approximate inflection-point behavior for the inflaton potential at $\phi = M$.

%%%%%%%%%%%%%%%%%%%%%%%%%%%%%%%%%%
\subsection{Thermal Leptogenesis and Reheating}
\label{sec:Reheat}
%%%%%%%%%%%%%%%%%%%%%%%%%%%%%%%%%%
Leptogenesis \cite{Fukugita:1986hr} is a relatively simple  mechanism to generate the observed baryon asymmetry in a model with the type-I seesaw mechanism. 
If Majorana neutrinos are non-degenerate in mass, 
a successful thermal leptogenesis scenario requires the lightest Majorana neutrino mass ($m_{N^1}$) to be heavier than $10^{9-10}$ GeV   
and the reheat temperature $T_R > m_{N^1}$ \cite{Buchmuller:2002rq}. 
In our setup, the $U(1)_X$ gauge interactions \cite{Iso:2010mv} and Yukawa interactions \cite{Dev:2017xry} of the Majorana neutrinos can keep these neutrinos in thermal equilibrium with the SM particles. 
These processes will suppress the generation of lepton asymmetry until they freeze out. 
By requiring these processes to decouple before the temperature of the thermal plasma drops to $T \sim m_{N^1}$, 
we now derive the conditions necessary to prevent such a suppression.

We first consider the $Z^\prime$ mediated process, 
${(N^c)^1} (N^c)^1 \to  Z^\prime \to \overline{f_{SM}} f_{SM} $, 
where $f_{SM}$ are the SM fermions.  
Since $m_{Z^\prime} > m_{N^{1}}$, 
the $Z^\prime$ mediated process is effectively a four-Fermi interaction, 
and its thermal-averaged cross section for $T \gtrsim m_{N^1}$ can be approximated as \cite{Okada:2016ssd}
\bea
\langle \sigma v\rangle \simeq \frac{F(x_H)}{768\pi} \frac{T^2}{v_X^4} , 
\eea 
where $F(x_H) = 13 + 16 x_H + 10x_H^2$. 
The process decouples at $T\sim m_{N^1}$, 
if $\left.\Gamma/H\right|_{T=m_{N^1}}< 1$, 
where 
$\Gamma (T) =  n_{eq}(T) \langle \sigma v\rangle $ 
is the annihilation/creation rate of $(N^c)^1$ with an equilibrium number density $n_{eq}(T) \simeq 2 T^3/\pi^2$,  
and 
$H(T) \simeq \pi T^2/M_P$ 
is the corresponding value of the Hubble parameter.     
This leads to a lower bound  
\bea
v_{X} > 2.17 \times 10^{10} \;{\rm GeV} \left( F(x_H)\right)^{1/4}\left(\frac{m_{N^1}}{10^{9} \; {\rm GeV}}\right)^{3/4}. 
\label{eq:cond1}
\eea

Because we require $M<0.11 M_P$ to solve the axion domain wall and isocurvature problems, 
it follows from Eq.~(\ref{eq:ratiophiz}) that $m_{N^1} > m_\phi$. 
In this case, another process, ${N_R}^{1,2} N_R^{1,2} \leftrightarrow  \phi \phi$, can suppress the generation of lepton asymmetry as investigated in Ref.~\cite{Dev:2017xry}. 
The thermal-averaged cross section of this process is roughly given by \cite{Plumacher:1996kc}
\bea
   \langle \sigma v \rangle \simeq \frac{1}{4 \pi}   \frac{{m_{N^1}}^2}{v_{BL}^4}. 
\label{eq:cs2}
\eea
Requiring $\Gamma/H < 1$ at $T=m_{N^1}$ to avoid the suppression of the generation of lepton asymmetry, we obtain  
\bea
v_{X} > 7.92 \times 10^{10} \;{\rm GeV} \left(\frac{m_{N^1}}{10^{9} \; {\rm GeV}}\right)^{3/4}.  
\label{eq:cond2}
\eea

For the remainder of this section, 
let us fix $x_H = 0$, $M = 0.05M_P$ and $m_{N^1} = 10^9$ GeV to be our benchmark values, and we find $v_X \simeq 1.93 \times 10^{12}$ GeV,  
$m_\phi \simeq 9.30 \times 10^4$ GeV and $m_{N^{2,3}} \simeq 9.32 \times 10^9$ GeV from Eq.~(\ref{eq:ratiophiz}).   
This value of $v_X$ is consistent with   Eq.~(\ref{eq:cond2}), which is stronger than condition in Eq.~(\ref{eq:cond1}) for $x_H = 0$. 
The observed baryon asymmetry therefore can be produced by thermal leptogenesis if the reheat temperature $T_R>m_{N^1}$.

%For $m_{N^1} = 10^{9}$ GeV and $10= m_{N^1}/m_{Z^\prime}$,  we obtain the following expression for the $U(1)_X$ VEV using Eq.~(\ref{eq:ratiophiz}), 
%\bea
%v_X \simeq 6.87\times 10^{11} \; {\rm GeV} \left(100 + 257 x_H + 165 x_H^2 \right)^{1/6}  \left(\frac{R}{10}\right)\left(\frac{0.11\; M_P}{M}\right)^{1/3}. 
%\eea

The SM particles produced from the decay of the inflation reheat the universe, 
thereby connecting inflation to the standard hot big bang cosmology.  
To estimate the reheat temperature, 
we assume an instantaneous inflaton decay, which yields the standard formula
\begin{eqnarray}
    T_R \simeq \left(\frac{90}{\pi^2 g_*}\right)^{1/4} \sqrt{\Gamma_\phi M_P},  
\label{eq:TR}
\end{eqnarray} 
where $g_* \simeq 100$ and $\Gamma_\phi$ is the total decay width of the inflaton. 
To estimate the inflaton decay width, 
we consider the following interactions between $\Phi$ and the SM doublet Higgs fields,   
\bea
V \supset  2 \lambda^\prime  \left(\Phi^\dagger \Phi\right) \left(H_u^\dagger H_u + H_d^\dagger H_d\right)
\supset 
 2 \lambda^\prime v_{X}  \phi \left(H_u^\dagger H_u + H_d^\dagger H_d\right),   
\label{eq:InfPot}
\eea 
The decay width of $\phi$ can be approximated as 
\bea
\Gamma_\phi \simeq  \frac{{\lambda^\prime}^2 v_X^2 }{\pi \, m_ \phi}, 
\label{eq:gamma2}
\eea
where we have neglected the mass of Higgs bosons in the final state. The reheat temperature is given by    
\bea
T_R \simeq 10^{10} \;{\rm GeV} \left(\frac{\lambda^\prime}{2.31 \times 10^{-9}}\right).  
\label{Lambda}
\eea 
Thus, an adequate reheat temperature for successful thermal leptogenesis, $T_R >m_{N^1}$ can be achieved with $\lambda^\prime \gtrsim 2.31\times10^{-9}$. 
Such a small value of $\lambda^\prime$ has a negligible effect on the effective potential of the inflaton and can be safely ignored as far as the inflationary phase is concerned.

%%%%%%%%%%%%%%%%%%%
\section{SU(5) Grand Unification and SMART U(1)$_X$}
\label{sec:}
%%%%%%%%%%%%%%%%%%%
In this section 
we discuss how the SMART U(1)$_X$ model can be merged with $SU(5)$ grand unification. 
In Table~\ref{tab:1}, for $x_H=-4/5$,  
the SM quarks and leptons  are unified in $SU(5)\times U(1)_X \times U(1)_{PQ}$ multiplets: 
${(F_{5}^*)}^i$ $({\bf 5^*}, -3/5, -1) \supset (d^c)^{i} \oplus \ell^{i}$ 
and 
$F_{10}^i$ of $({\bf 10}, 1/5, 1) \supset q^{i} \oplus (d^c)^{i} \oplus  (e^c)^{i}$. 
The Higgs multiplets in ${\bf 5}$ and ${\bf 5}^*$ representation of $SU(5)$ contain 
the SM doublet Higgs fields $H_{u,d}$, 
namely, $({\bf 2}, +1/2,-2) \supset H_u$ and  $({\bf 2}, -1/2,-2) \supset H_d$.  
A simple scenario realizing the unification of the SM gauge couplings at around $M_{GUT} \simeq 4.0 \times 10^{16}$ can be achieved in the presence of a pair of vector-like quarks with mass of ${\cal O}$(TeV) \cite{GCU1, GCU2}. 
To implement this, 
we include two sets of new vector-like fermions, namely,  
$V_5+{V_5}^*=({\bf 5}, 3/5, 1)+({\bf 5^*}, -3/5, -1)$ 
and 
$V_{10}+{V_{10}}^*=({\bf 10}, 1/5, 1)+({\bf 10^*}, -1/5, -1)$. 
Furthermore, a $U(1)_X$ and $U(1)_{PQ}$ charge neutral Higgs field in the adjoint representation of $SU(5)$, 
which breaks $SU(5)$ symmetry to the SM, is used to generate a mass-splitting between the triplet and doublet components of the new fermions \cite{Okada:2017dqs}. 
We fix the model parameters such that only a pair of vector-like quarks, 
$D+D^c$ and $Q+Q^c$, 
remain light at low energies whose representation under the SM gauge group is the same as $(d^c)^{i}$ and $q^{i}$, respectively.

The presence of the new quarks can also stabilize the SM Higgs potential\footnote{See Refs.~\cite{Basler:2017nzu} for a detailed study of the stability of the two Higgs doublet potential with the inclusion of all the mixed quartic coupling terms in the potential.} at high energies \cite{GCU2}. 
We shall work in the so-called alignment limit for the two Higgs doublet model with $\tan \beta = v_u/v_d \gg 1$, 
such that 
$H_u \supset h \sin\beta $ and the quartic self-coupling $\lambda_u$ can be effectively identified with the SM Higgs (h) and its  quartic self-coupling ($\lambda_h$), respectively \cite{Branco:2011iw}. 
We fixed the mixed quartic coupling between $H_u$ and $H_d$ to be negligible in our analysis. 
Following Ref.~\cite{GCU2}, we numerically solve the RGEs. 
We set two vector-like quark pairs and the new Higgs doublet mass to be  2.5 TeV.

\begin{figure}[t]
\begin{center}
\includegraphics[scale=0.8]{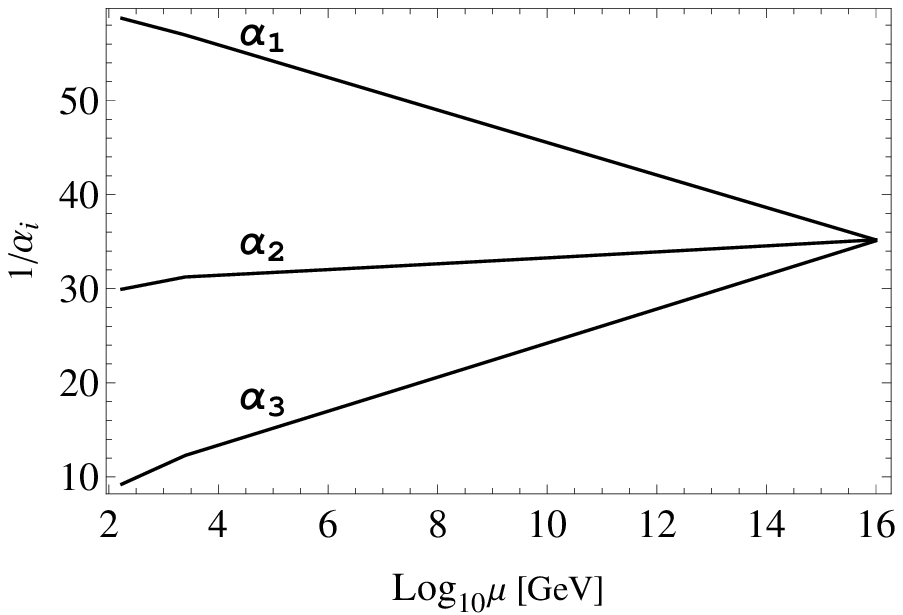} \;
\includegraphics[scale=0.6]{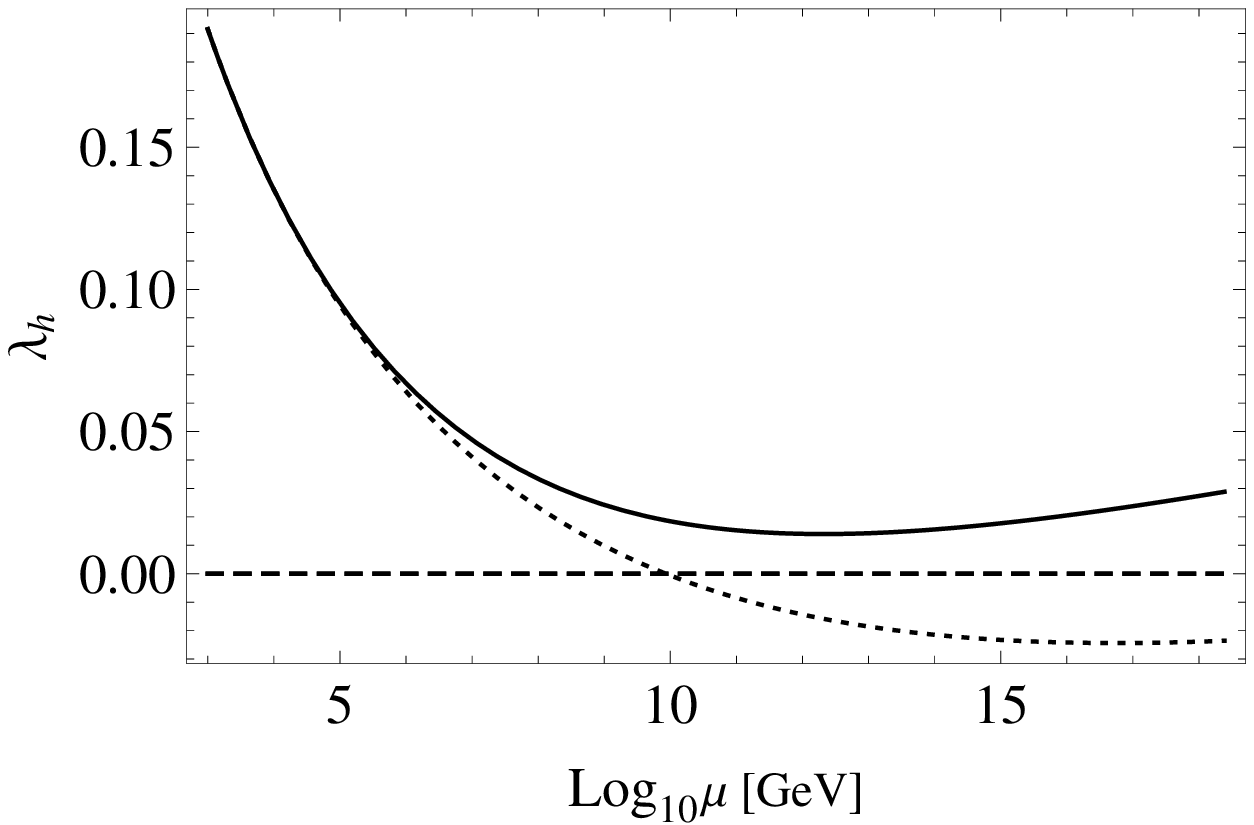}       
\end{center}
\caption{
Renormalization group running of the SM couplings including vector-like quarks with degenerate masses of 1 TeV. Top panel: the diagonal solid lines labeled $\alpha_i = g_i^2/4\pi$ ($i=1,2,3$) depict the SM $U(1)_Y$, $SU(2)_L$ and $SU(3)_c$ gauge coupling, respectively, which are unified at $M_{\rm GUT} \simeq 9.8 \times 10^{15}$ GeV. 
Bottom panel: the solid (dotted) curve depicts the RG running of the SM Higgs quartic coupling with (without) the inclusion of the new vector-like quarks, and the horizontal dashed line depicts $\lambda_h = 0$.  
}
\label{fig:SU5}
\end{figure}

In the left panel of Fig.~\ref{fig:SU5}, 
we plot the RG running of the SM gauge couplings as a function of the energy scale $\mu$. 
The diagonal solid lines labeled $\alpha_i = g_i^2/4\pi$ ($i=1,2,3$) depict the SM gauge couplings for $U(1)_Y$, $SU(2)_L$ and $SU(3)_c$, respectively. 
The SM gauge couplings successfully unify at around $M_{\rm GUT} \simeq 9.8 \times 10^{15}$ GeV, with $\alpha_{GUT} \simeq 1/35$. 
With these values, 
the gauge boson mediated proton lifetime\footnote{
To prevent rapid proton decay, the color triplet Higgs field contained in Higgs fields in the ${\bf 5}$ and ${\bf 5}^*$ representations
of $SU(5)$ must have mass greater than $10^{13}$ GeV \cite{Nath:2006ut}. 
We simply set their masses to be $M_{GUT}$.
%and we have ignored their effect on running of the SM gauge couplings.
} is estimated to be \cite{Nath:2006ut}  
\bea
\tau_p \approx \frac{1}{\alpha_{GUT}^2} \frac{M_{GUT}^4}{m_p^5} \approx 2.6 \times 10^{35} \; {\rm years}, 
\eea  
where $m_p = 0.983$ GeV is the proton mass. 
This is consistent with the experimental lower bound obtained by the Super-Kamiokande, 
$\tau_p(p \to \pi^0 e^+) \gtrsim 10^{34}$ yr \cite{Miura:2016krn}. 
The predicted lifetime is close to the sensitivity limit of future Hyper-Kamiokande, $\tau_p \lesssim 1.3 \times 10^{35}$ yr \cite{Abe:2011ts}.

In the right panel, the solid (dotted) curve depicts the RG running of the SM Higgs quartic coupling with (without) the new vector-like quarks and a the new Higgs doublet contributions, and the horizontal dashed line depicts $\lambda_h = 0$. 
In the presence of the new vector-like fields, $\lambda _h(\mu) > 0$, and thus the SM Higgs potential is stabilized.

%%%%%%%%%%%%%%%%%%%%%%%%%%%%%%%%
%%%%%%%%%%%%%%%%%%%%%%%%%%%%%%%%%%

Next we study the IPI scenario. 
Note that in the SMART U(1)$_X$ case and we have $\lambda_\phi(M) < 0$ for $-0.87 \lesssim x_H \lesssim-0.69$ from Eq.~(\ref{eq:LandG}), which implies that the inflaton potential is unstable. 
With the inclusion of the new vector-like quarks ($D+D^c$ and $Q+Q^c$), 
which is essential for the unification of the SM gauge couplings, 
we now show that the inflaton potential remains stable for any value of $x_H$.  
As in the SMART U(1)$_X$ case, 
we fix $Y_1 (M)/g(M) = \sqrt{2}/5$ and $Y_{2,3} (M)= Y(M)$, 
such that the conditions $\beta_{\lambda_\phi} (M) = 0$ leads to $Y(M) \simeq 2.63\; g(M)$, which is the same relation as before.  
However, in the presence of the new vector-like quarks and the extra Higgs doublet, 
which are also charged under the  $U(1)_X$ symmetry, 
the $U(1)_X$ gauge coupling RG equation in Eq.~(\ref{eq:RGEs}) is modified to  
\bea
 \phi  \frac{d g}{d  \phi} &=& \frac{1}{16 \pi^2}  \left(\frac{40 + 32 x_H + 23 x_H^2}{6}\right) g^3.     
\label{eq:RGEsSU5}
\eea 
This leads to a new relation between the quartic and gauge couplings at the inflation scale $\phi = M$,   
\bea
\lambda_\phi(M) \!\simeq\! 5.23\times 10^{-5}\! \left(100+196x_H+141 {x_H}^2\right)\! g(M)^6.  
\label{eq:FEqLSU5}
\eea
which is positive for any value of $x_H$. 
Therefore the inflaton potential is stable and we can realize a successful IPI scenario for $x_H = -4/5$.

To discuss reheating and leptogenesis, 
we next consider the low energy predictions for the masses and couplings. 
Since the quartic coupling at low energies only depends on $\lambda_\phi (M)$, 
its low energy value is still determined by Eq.~(\ref{eq:Lmu}). 
Comparing Eq.~(\ref{eq:FEqLSU5}) with $\lambda_\phi (M)$ value given by Eq.~(\ref{eq:FEq}), 
the $U(1)_X$ gauge coupling at the inflation scale $\phi = M$ is given by 
\bea
g(M)\simeq  \frac{1.45\times 10^{-3}}{\left(100+196x_H+141 {x_H}^2\right)^{1/6}}  
 \;\left(\frac{M}{M_{P}}\right)^{1/3}.  
\label{eq:SU5FEq2} 
\eea
The low energy values of the gauge and Yukawa couplings are well approximated by their values at $\phi=M$, 
$g(v_X) \simeq g(M) $ and $  Y_i (v_X) \simeq Y_i (M)$. 
For a benchmark value $M= 0.05 M_P$, with the lightest Majorana neutrino mass $m_{N^1} = 10^{10} \; {\rm GeV}= m_{Z^\prime}/10$, 
we obtain $v_X \simeq 1.68 \times 10^{12}$ GeV,  
and the particle masses are given by $m_\phi \simeq 8.23 \times 10^4$ GeV, 
and $m_{N^{2,3}} \simeq 9.30 \times 10^9$ GeV. 
The conditions for preventing a suppression of lepton asymmetry from the gauge (Yukwa) interactions of the lightest Majorana neutrino leads to a lower bound of $v_X > 2.46\;(7.92) \times 10^{10} $ GeV. 
This is consistent with the $v_X$ value obtained for the benchmark.  
The observed baryon asymmetry can be generated by thermal leptogenesis 
for $T_R \simeq 10^{10} > m_{N^1}$ GeV and this can be realized with $\lambda^\prime \gtrsim 2.31\times 10^{-6}$ as shown in Eq.~(\ref{Lambda}).

%%%%%%%%%%%%%%%%%
\section{Summary}
\label{sec:conc}
%%%%%%%%%%%%%%%%%
%Variety of cosmological and particle physics observations have demonstrated that 
%the Standard Model (SM) is at best an effective theory description of nature and needs to be supplemented with new physics beyond the SM to addresses its five fundamental shortcomings.  
%To achieve this goal, we have proposed the SMART U(1)$_X$ model:  SM with Axion, Right handed neutrinos, Two Higgs doublets and a gauged U(1) extension. 

We have proposed a SMART U(1)$_X$ model which is an anomaly free $U(1)_X $   extension of the SM supplemented by a $U(1)_{PQ}$ symmetry. 
The $U(1)_X$ charge of each particle is defined as a linear combination of its hypercharge and $B-L$ charge and is determined by a single free parameter $x_H$. 
Three right handed neutrinos (RNHs) are added to cancel all $U(1)_X$ related anomalies. 
These neutrinos, as is well-known, also explain the origin of the observed neutrino oscillations via the type-I seesaw mechanism and 
generate the observed baryon asymmetry via leptogenesis. 
The $U(1)_{PQ}$ symmetry solves the strong CP problem  
and also provides the axion as a compelling dark matter (DM) candidate. 
By identifying a $U(1)_X$ breaking Higgs field as the inflaton, 
we have implemented a low-scale inflection-point inflation (IPI) scenario in our model to realize a Hubble parameter during inflation $H_{inf} \lesssim 2 \times 10^{7}$ GeV. 
The $U(1)_X$ gauge symmetry is crucial for the implementation of the IPI scenario. 
This low-scale inflation is a key for resolving the axion domain wall and the axion DM isocurvature problems. 
After the end of inflation, 
we have shown that the inflaton decay adequately reheats the universe to allow for a successful implementation of the leptogenesis scenario.

We have also examined another possibility to merge the SMART U(1)$_X$ with $SU(5)$ grand unification. 
For $x_H=-4/5$,  
all the SM quarks and leptons are embedded within the  $SU(5)\times U(1)_X \times U(1)_{PQ}$ representations. 
We extended the particle content of the SMART U(1)$_X$ with a pair of new vector-like fermions and a new Higgs doublet with ${\cal O}(1)$ TeV mass to realize successful unification of the three SM gauge couplings at $M_{GUT}=9.8 \times 10^{15}$ GeV. 
This leads to proton lifetime estimate of $2.6 \times 10^{35}$ yr, which is close to the sensitivity limit of the future Hyper-Kamiokande experiment. 
These new fermions also help stabilize the SM Higgs potential as well as the inflaton potential. 
The latter is crucial for successful implementation of the IPI scenario which is a key for solving the two cosmological problems encountered in the axion DM scenario. 
The leptogenesis scenario is also implementable in the $SU(5)$ model.

%Finally, 
%we briefly show that our model can be compatible with the recently proposed Trans-Planckian Censorship Conjecture  
%which requires the Hubble parameter during the inflation to be relatively small, $H_{inf} < 1$ GeV. 
%This can be realized in the IPI scenario. 
%We have ro resort to resonant leptogenesis in order to explain the observed baryon asymmetry. 

%%%%%%%%%%%%%%%%%%%%%%%%%%%%%%%%%%%%%%%%%
\section*{Acknowledgements}
%%%%%%%%%%%%%%%%%%%%%%%%%%%%%%%%%%%%%%%%%
This work is supported in part by the United States Department of Energy grant DE-SC0012447 (N.~Okada) 
and DE-SC0013880 (D.~Raut and Q.~Shafi).

%%%%%%%%%%%%%%

\end{document}